\documentclass[11pt,a4paper]{article}
\usepackage{jheppub,amsmath,amssymb,slashed,url,bm,textgreek,upgreek}
\usepackage{graphicx}
\usepackage{epstopdf}

\newcommand\be{\begin{equation}}
\newcommand\ee{\end{equation}}
\newcommand\bea{\begin{eqnarray}}
\newcommand\eea{\end{eqnarray}}
\newcommand\bsp{\begin{split}}
\newcommand\esp{\end{split}}
\newcommand\bal{\begin{aligned}}
\newcommand\eal{\end{aligned}}

\newcommand{\rt}{\vcenter{\hbox{\includegraphics[scale=0.3]{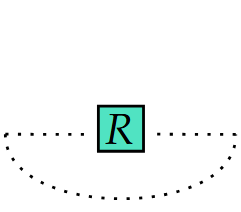}}}} 
\newcommand{\rtz}{\vcenter{\hbox{\includegraphics[scale=0.3]{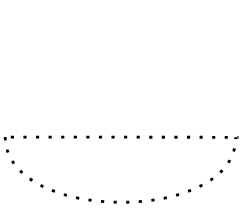}}}}
\newcommand{\rto}{\vcenter{\hbox{\includegraphics[scale=0.3]{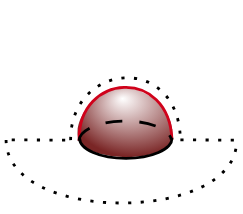}}}}
\newcommand{\rttd}{\vcenter{\hbox{\includegraphics[scale=0.3]{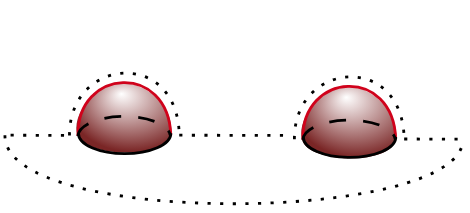}}}}
\newcommand{\rttc}{\vcenter{\hbox{\includegraphics[scale=0.3]{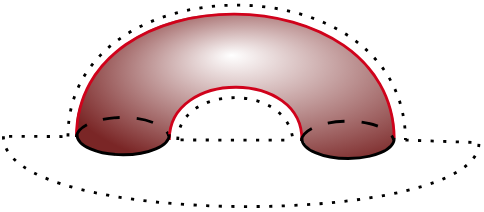}}}}
\newcommand{\srto}{\vcenter{\hbox{\includegraphics[scale=0.3]{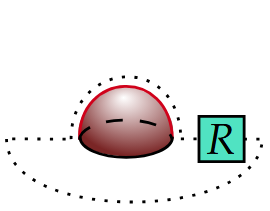}}}}
\newcommand{\srtt}{\vcenter{\hbox{\includegraphics[scale=0.3]{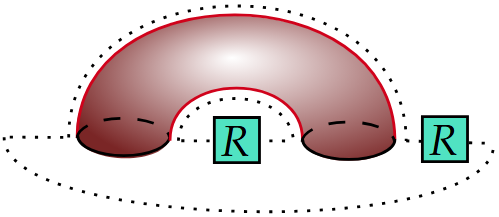}}}}

\newcommand{\FB}{\vcenter{\hbox{\includegraphics[scale=0.1]{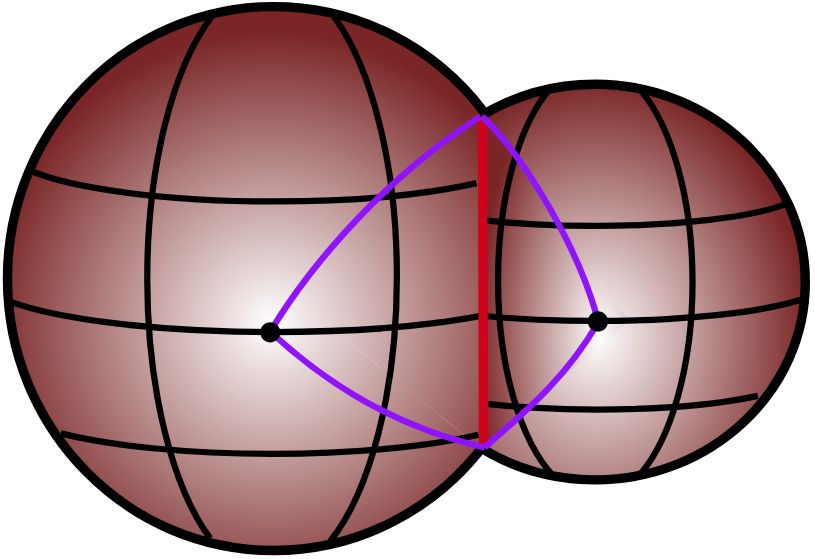}}}} 
\newcommand{\LL}{\vcenter{\hbox{\includegraphics[scale=0.1]{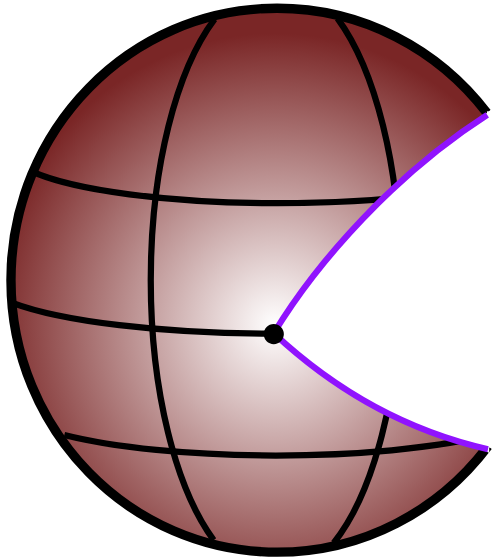}}}}
\newcommand{\LR}{\vcenter{\hbox{\includegraphics[scale=0.1]{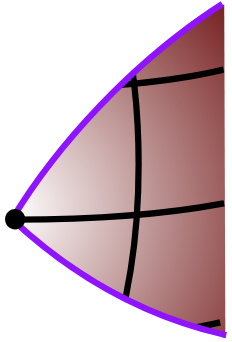}}}}
\newcommand{\BR}{\vcenter{\hbox{\includegraphics[scale=0.1]{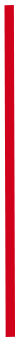}}}}
\newcommand{\RL}{\vcenter{\hbox{\includegraphics[scale=0.1]{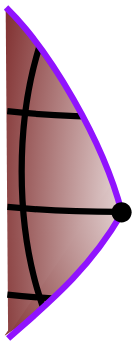}}}}
\newcommand{\RR}{\vcenter{\hbox{\includegraphics[scale=0.1]{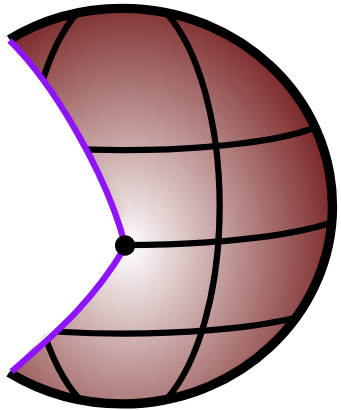}}}}

\newcommand{\FBs}{\vcenter{\hbox{\includegraphics[scale=0.03]{figures/FB.png}}}} 
\newcommand{\LLs}{\vcenter{\hbox{\includegraphics[scale=0.02]{figures/LL.png}}}}
\newcommand{\LRs}{\vcenter{\hbox{\includegraphics[scale=0.04]{figures/LR.png}}}}
\newcommand{\BRs}{\vcenter{\hbox{\includegraphics[scale=0.04]{figures/BR.png}}}}
\newcommand{\RLs}{\vcenter{\hbox{\includegraphics[scale=0.04]{figures/RL.png}}}}
\newcommand{\RRs}{\vcenter{\hbox{\includegraphics[scale=0.04]{figures/RR.png}}}}

\makeatletter 
\def\@fpheader{{\color{white}jhep}}
\title{\boldmath Microscopic Origin of the Entropy of De Sitter Spacetime}%
\author{Zhi Wang}
\affiliation{College of Physics, Nanjing University of Aeronautics and Astronautics, \\ Nanjing, 211106, China}
\emailAdd{zhiwang@nuaa.edu.cn}
\abstract{
We construct an infinite family of semiclassical de Sitter (dS) microstates, realized as backreacted geometries of dS spacetime with a constant tension thin-shell brane located outside the dS event horizon. We further show that wormhole contributions to the semiclassical Euclidean gravitational path integral lead to universal nonperturbative overlaps between these microstates. By evaluating the nonperturbative overlaps, we count the dimension of the Hilbert space spanned by the semiclassical dS microstates and find that it precisely equals the exponential of the Gibbons-Hawking entropy of dS spacetime. Our construction thus provides a state-counting derivation for Gibbons-Hawking entropy of dS spacetime.
}

\begin{document} \maketitle

\section{Introduction}
Black holes and cosmology are two primary arenas for exploring quantum aspects of gravity. Not long after Bekenstein and Hawking~\cite{Bekenstein:1973ur,Hawking:1975vcx} discovered that an entropy should be associated with the area of a black hole horizon, Gibbons and Hawking~\cite{Gibbons:1977mu} showed that the area of a cosmological horizon, divided by $4G_N$, should also be interpreted as an entropy. For an ordinary macroscopic quantum system, thermodynamic entropy originates microscopically from the underlying quantum states and equals the logarithm of the dimension of its Hilbert space. However, due to the lack of a complete quantum mechanical description for gravity in black hole and cosmological spacetimes, understanding the microscopic origin of their entropy remains a fundamental challenge.

Recent advances in the Euclidean path integral approach to quantum gravity have shed new light on this problem. By incorporating nonperturbative wormhole contributions, the semiclassical Euclidean gravitational path integral has proven useful for extracting fine-grained information about the underlying quantum gravity theory. In particular, it has been used to derive the Page curve for the fine-grained entropy of Hawking radiation in an evaporating black hole~\cite{Penington:2019kki,Almheiri:2019qdq,Almheiri:2020cfm}. Moreover, it provides a method for computing nonperturbative overlaps between semiclassical black hole microstates, which has led to a state-counting derivation for Bekenstein-Hawking entropy of black holes in asymptotically anti-de Sitter or flat spacetimes~\cite{Balasubramanian:2022gmo,Balasubramanian:2022lnw} (see also~\cite{Climent:2024trz,Balasubramanian:2024yxk,Geng:2024jmm,Banerjee:2024fmh,Balasubramanian:2024rek,Balasubramanian:2025jeu,Balasubramanian:2025zey,Balasubramanian:2025hns}). These developments in the understanding of black hole entropy motivate us to further investigate the microscopic origin of the entropy of cosmological spacetimes.

Observational evidence indicates that our universe underwent an early inflationary phase~\cite{Planck:2018jri} and is currently undergoing a late-time accelerated expansion~\cite{SupernovaSearchTeam:1998fmf}, both of which are locally approximated by a dS spacetime. However, the Gibbons-Hawking entropy of dS spacetime is more mysterious than black hole entropy. One conceptual problem arises from the fact that the spatial geometry of dS spacetime is closed. This implies the absence of a timelike boundary to directly define the thermal ensemble and thermodynamic quantities for the dS gravitational system, let alone understanding the microscopic origin of its entropy~\footnote{Ref.~\cite{Coleman:2021nor} provided a holographic understanding for the entropy of dS$_3$ by connecting dS$_3$ gravity with $T\bar T\left(+\Lambda_2\right)$ deformation of AdS/CFT.}. A useful approach to overcoming this problem is to view dS thermodynamics as the vanishing-wall limit of the quasi-local thermodynamics of dS gravity with a timelike Dirichlet wall~\cite{Banihashemi:2022jys,Banihashemi:2022htw}. The introduction of a Dirichlet wall to dS spacetime provides a thermal ensemble interpretation for dS entropy and enables the construction of various semiclassical dS microstates. In this work, we construct an infinite family of semiclassical dS microstates by considering the dS geometry with a Dirichlet wall, backreacted by a constant tension thin-shell brane located outside the dS event horizon. Furthermore, we employ these semiclassical dS microstates to provide a gravitational state-counting derivation for the Gibbons-Hawking entropy of dS spacetime. Nonperturbative wormhole contributions in semiclassical Euclidean gravitational path integral play a central role in this state-counting derivation.

\section{Gibbons-Hawking entropy of dS spacetime}
\begin{figure}
    \centering
\includegraphics[width=0.23\linewidth]{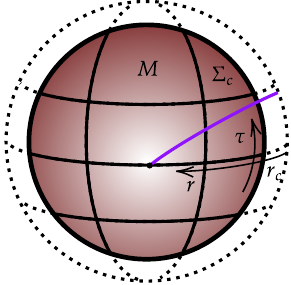}
\quad\quad\quad\quad\quad\quad\quad\quad
\includegraphics[width=0.23\linewidth]{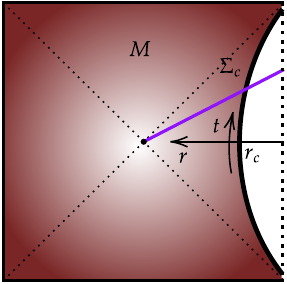}
    \caption{Left: Euclidean dS$_{d+1}$ saddle point geometry with a Dirichlet wall at $r = r_c$. Right: Penrose diagram of the corresponding Lorentzian spacetime obtained by Wick rotation. Each point in the figures represents a $\left(d-1\right)$-dimensional sphere of radius $r$. }
    \label{dS wall}
\end{figure}

Consider the $(d+1)$-dimensional Einstein gravity theory with a positive cosmological constant $\Lambda$ and a timelike Dirichlet wall $\Sigma_c$, described by the action
\be\label{lac}
I_{M_c}=\frac{1}{16\pi G_N}\int_{M}\sqrt{-g}\,(R-2\Lambda)+\frac{1}{8\pi G_N}\int_{\Sigma_c}\sqrt{-\gamma}\,K_c\ ,
\ee
where the Gibbons-Hawking-York boundary term is added on the Dirichlet wall. The canonical (or microcanonical) ensemble of this gravitational system is defined by fixing the quasi-local temperature (or energy) along the Dirichlet wall. The quasi-local thermodynamics of this gravitational system can then be derived from the corresponding thermal ensemble using the Euclidean approach~\cite{Gibbons:1976ue}.
In the semiclassical limit with $G_N\to 0$, the path integral representation of the thermal partition function for the quasi-local thermodynamics is dominated by the Euclidean saddle point geometries that satisfy the specified quasi-local boundary conditions. In particular, under the spherical symmetry reduction, Euclidean pure dS geometry with a cosmological horizon enclosed by the Dirichlet wall (Fig.~\ref{dS wall}) is a stable dominant saddle of the canonical partition function (see~\cite{Banihashemi:2022jys} for the phase diagram). 

The metric of the Euclidean pure dS geometry with a Dirichlet wall at $r=r_c$ is
\be\label{pds}
\mathrm{d}s^2=f(r)\mathrm{d}\tau^2+\frac{1}{f(r)}\mathrm{d}r^2+r^2\mathrm{d}\Omega_{d-1}^2\ ,\quad f(r)=1-\frac{r^2}{l^2}\ ,
\ee
where $l=\sqrt{\frac{d(d-1)}{2\Lambda}}$ is the dS radius and $r\in[r_c,l]$ describes the region in the interior of the Dirichlet wall. Under the Wick rotation $t=-i\tau$, it corresponds to the static patch of dS spacetime with a Dirichlet wall at $r=r_c$ (Fig.~\ref{dS wall}). The quasi-local temperature and quasi-local energy associated with the Dirichlet wall are given by~\cite{Banihashemi:2022htw}
\be
\bal
T_c=\frac{1}{\beta_c}=\frac{1}{2\pi l}\frac{1}{\sqrt{1-\frac{r_c^2}{l^2}}}\ ,\quad
E_c=\frac{\left(d-1\right)r_c^{d-2}}{8\pi G_N}\sqrt{1-\frac{r_c^2}{l^2}}V_{\mathrm{S}^{d-1}}\ ,
\eal
\ee
where $V_{\mathrm{S}^{d-1}}=\frac{2\pi^{\frac{d}{2}}}{\Gamma\left(\frac{d}{2}\right)}$ is the volume of the $(d-1)$-dimensional unit sphere. 
The canonical partition function dominanted by the Euclidean pure dS geometry can be written as
\be\label{onshell}
Z(\beta_c)=\int Dg e^{-I^E_{M_c}[g]}\approx e^{-I^E_{M_c}}=e^{-\beta_c (E_c-T_c S)}\ , 
\ee
where the Euclidean on-shell action is given by
\be
I^E_{M_c}=\frac{\left(d-1\right)r_c^{d-2}\left(l^2-r_c^2\right)-l^d}{4G_Nl}V_{\mathrm{S}^{d-1}}\ .
\ee
Then the canonical entropy of the dS static patch with a Dirichlet wall at $r=r_c$ is derived from (\ref{onshell}) as~\footnote{The entropy can be derived in the same way from the microcanonical ensemble by fixing the quasi-local energy~\cite{Banihashemi:2022jys}.}
\be\label{ent}
\bal
S&=\beta_cE_c-I^E_{M_c}=\frac{l^{d-1}V_{\mathrm{S}^{d-1}}}{4G_N}=\frac{A}{4G_N}\ ,
\eal
\ee
where $A$ is the area of the dS horizon. 

The entropy in (\ref{ent}) is independent of the position of the Dirichlet wall and corresponds to the Gibbons-Hawking entropy of the dS horizon. Moreover, as explained in~\cite{Banihashemi:2022jys}, in the vanishing limit of the Dirichlet wall with $r_c \rightarrow 0$, the quasi-local thermodynamics of the dS static patch with a Dirichlet wall reproduces the thermodynamics of the full dS static patch. Thus, the above construction provides a thermal ensemble interpretation for the Gibbons-Hawking entropy of dS spacetime. It also offers a framework for further investigation of its microscopic origin.

\section{Semiclassical dS microstates}
A macrostate of the dS static patch, defined by fixed values of thermodynamic quantities along the Dirichlet wall, corresponds to many microstates in the fundamental Hilbert space of the dS gravity theory. Semiclassically, some of these microstates can be realized as dS gravitational solutions distinguished by the geometry outside the dS horizon. We now construct an infinite family of such semiclassical dS microstates, realized as backreacted geometries of dS spacetime with a constant tension thin-shell brane located outside the dS horizon.

\begin{figure}
\centering
 \includegraphics[width=0.45\linewidth]{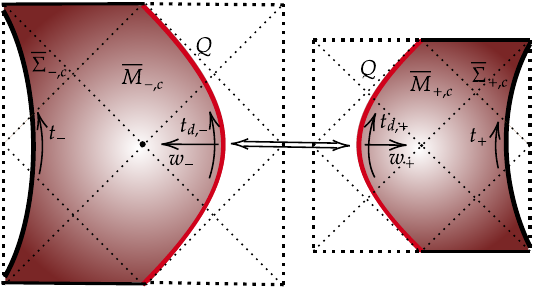}
    \caption{Penrose diagram of two dS$_{d+1}$ spacetimes glued along a dS$_d$ thin-shell brane.}
    \label{dSL}
\end{figure}

The action of dS gravity bounded by a timelike Dirichlet wall and backreacted by a constant tension thin-shell brane can be written as
\be
\bal\label{lpmac}
I_{\bar{M}_c}&=I_{\bar{M}_{-,c}}+I_{Q}+I_{\bar{M}_{+,c}}\ ,\\
I_{\bar{M}_{\pm,c}}&=\frac{1}{16\pi G_N}\int_{\bar{M}_{\pm}}\sqrt{-g_\pm}\,(R_{\pm}-2\Lambda_\pm)+\frac{1}{8\pi G_N}\int_{\bar{\Sigma}_{\pm,c}}\sqrt{-\gamma_{\pm}}\,K_{\pm,c}\ ,\\
I_{Q}&=\frac{1}{8\pi G_N}\int_{Q}\sqrt{-h}\left(K_{+,Q}+K_{-,Q}+\left(d-1\right)T\right)\ .
\eal
\ee
Here, we consider a two-sided backreacted geometry $\bar{M}_c$, which consists of two bulk spacetimes, $\bar{M}_{-,c}$ and $\bar{M}_{+,c}$, glued along the thin-shell brane $Q$. In the action of the thin-shell brane $Q$, $K_{\pm,Q}$ are traces of the extrinsic curvature of $Q$ as embedded in $\bar{M}_{\pm,c}$, and $T$ is a free tension parameter of $Q$~\footnote{We treat $T$ as a cosmological constant term on the brane, which can take either positive or negative values~\cite{Randall:1999vf}. The physical tension of the thin-shell brane is $-T$~\cite{Marvel:2008uh}.}. We are interested in solutions of (\ref{lpmac}) where the two bulk spacetimes $\bar{M}_{\pm,c}$ are pure dS spacetimes of radii $l_{\pm}$, each containing a Dirichlet wall $\Sigma_{\pm,c}$ at $r = r_c$, and are glued along a spherical thin-shell brane $Q$ located outside the dS horizons (Fig~\ref{dSL}).  
The configuration of the spherical thin-shell brane $Q$, as well as the resulting glued geometry, is determined by the Israel junction conditions~\cite{Israel:1966rt}.

The trajectory of the spherical thin-shell brane embedded in $\bar{M}_{\pm,c}$ can be parametrized by its proper time $t_b$ as $\left(t_{\pm}(t_b),R(t_b)\right)$. The equation of motion of the brane trajectory is derived from the Israel junction conditions as (see appendix~\ref{appA})
\be\label{eomb}
\left(\frac{\mathrm{d}R}{\mathrm{d}t_b}\right)^2+V_{\mathrm{eff}}(R)=0\ ,\quad V_{\mathrm{eff}}(R)=1-\frac{R^2}{l_b^2}\ ,
\ee
where 
\be
l_b=\frac{2Tl_+^2l_-^2}{\sqrt{\left(l_+^2+l_-^2+l_+^2l_-^2T^2\right)^2-4l_+^2l_-^2}}
\ee
is the radius of the brane at the turning point of its trajectory. From Eq.~(\ref{eomb}), the trajectory of the thin-shell brane can be analytically obtained as
\be\label{rtb}
R(t_b)=l_b\cosh\frac{t_b}{l_b}\ ,\quad t_{\pm}(t_b)=-l_{\pm}\mathrm{arctanh}\frac{l_b\sinh\frac{t_b}{l_b}}{\sqrt{l_\pm^2-l_b^2}}\ .
\ee

\begin{figure}
\centering
\includegraphics[width=0.45\linewidth]{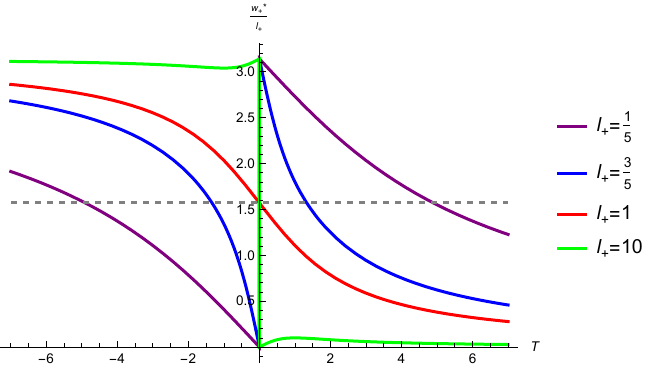}\quad\quad
\includegraphics[width=0.45\linewidth]{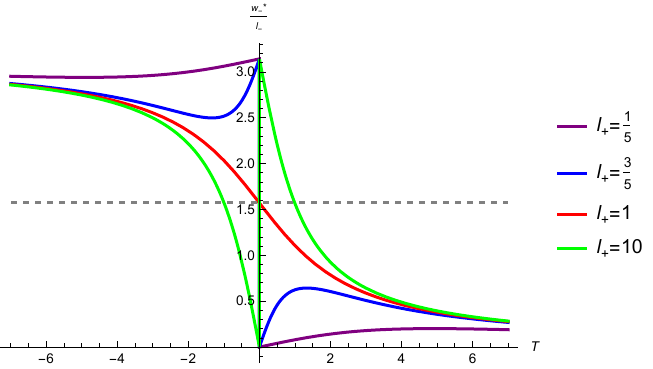}
    \caption{A plot of $\frac{w_{\pm}^*}{l_{\pm}}$ as a function of $T$ is shown for different values of $l_+$, where without loss of generality, we have set $l_{-}=1$.}
    \label{wT}
\end{figure}

Note that the dS static patch coordinates only describe the portion of the brane with $l_b\leq R(t_b)\leq l_{\pm}$. However, we can see that the physical radius of the brane in Eq.~(\ref{rtb}) coincides with the warp factor of a dS spacetime. This implies that the brane configuration is a dS$_d$ slice in each of the dS$_{d+1}$ spacetimes~\footnote{The composed dS geometry resembles that of two AdS$_{d+1}$ spacetimes glued along an AdS$_d$ brane, which has been proposed as the bulk dual of the holographic interface CFT~\cite{Simidzija:2020ukv}.}. A more natural coordinates to describe the entire brane trajectory is the dS/dS slicing coordinate~\footnote{The transformations between the static-patch coordinates and the dS/dS slicing coordinates are given by: $\sqrt{1-\frac{r^2}{l_\pm}}\sinh\frac{t_{{\pm}}}{l_{\pm}}=\sin\frac{w_{\pm}}{l_{\pm}}\sinh\frac{t_{d,{\pm}}}{l_{\pm}},\,\sqrt{1-\frac{r^2}{l_\pm}}\cosh\frac{t_{{\pm}}}{l_{\pm}}=\cos\frac{w_{\pm}}{l_{\pm}}$.} (Fig.~\ref{dSL})
\be
\mathrm{d}s_{\pm}^2=\mathrm{d}w_{\pm}^2+\sin^2 \frac{w_{\pm}}{l_{\pm}}\left(-\mathrm{d}t_{d,{\pm}}^2+l_{\pm}^2\cosh^2 \frac{t_{d,{\pm}}}{l_{\pm}} \mathrm{d}\Omega_{d-1}^2\right)\ .
\ee 
In these coordinates, the brane trajectory is specified by
\be
w_{\pm}=w_{\pm}^*=l_{\pm}\arcsin\frac{l_b}{l_{\pm}}\ . 
\ee
A plot of $w_{\pm}^{*}$ as a function of $T$ and $l_{\pm}$ is shown in Fig.~\ref{wT}.

\begin{figure}
\centering
\includegraphics[width=0.38\linewidth]{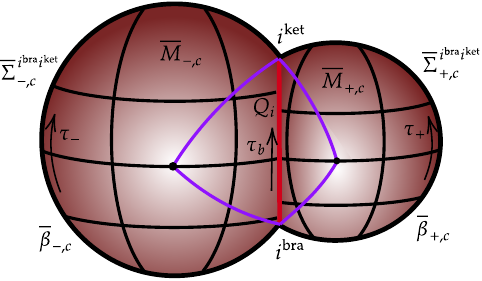}\quad\quad\quad\quad\quad\quad\quad
\raisebox{0.5cm}{\includegraphics[width=0.32\linewidth]{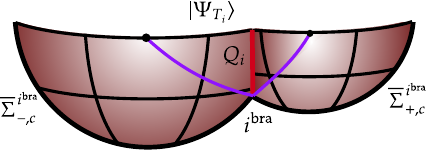}}
    \caption{Left: The Euclidean geometry of the dS$_{d+1}$ solution with both a Dirichlet wall and a thin-shell brane. Right: The Euclidean gravitational path integral in the lower half of the Euclidean dS geometry defines a semiclassical dS microstate on the time-reflection symmetric surface. }
    \label{dS}
\end{figure}

The dS gravitational solutions constructed above are symmetric under time reflection about the $t_{\pm}=0$ surface, and the Wick rotation of these solutions yields a Euclidean dS geometry with a regular Dirichlet wall $\bar{\Sigma}_c\equiv \bar{\Sigma}_{-,c}\cup\bar{\Sigma}_{+,c}$ as its boundary. In the Euclidean geometry, the thin-shell brane at $w_{\pm}=w^*_{\pm}$ lies along a line of constant latitude on the $\left(\tau_{\pm},r\right)$ hemisphere (Fig.~\ref{dS}). The Euclidean dS geometries backreacted by the thin-shell branes $Q_i$ with different tension parameters $T_i$ correspond to different Dirichlet walls $\bar{\Sigma}^{i^{\mathrm{bra}}i^{\mathrm{ket}}}_c\equiv\bar{\Sigma}_{-,c}^{i^{\mathrm{bra}}i^{\mathrm{ket}}}\cup\bar{\Sigma}_{+,c}^{i^{\mathrm{bra}}i^{\mathrm{ket}}}$, where we label the boundary conditions at the past and future intersection points between the Dirichlet wall and the thin-shell brane with tension parameter $T_i$ as $i^{\mathrm{bra}}$ and $i^{\mathrm{ket}}$, respectively. The proper length of the Dirichlet walls $\bar{\Sigma}^{i^{\mathrm{bra}}i^{\mathrm{ket}}}_c$ in the Euclidean dS geometry is given by (see appendix~\ref{appA})
\be\label{barb}
\bar{\beta}^i_{\pm,c}=\beta_{\pm,c}-\sqrt{1-\frac{r_{c}^2}{l_{\pm}^2}} \Delta\tau^i_{\pm}\ ,
\ee
where $\Delta \tau^i_{\pm}=2l_{\pm}\arctan \sqrt{\frac{l_{b}^2-r_c^2}{l_{\pm}^2-l_b^2}}$ are Euclidean boundary time elapsed by the thin-shell brane $Q_i$. 

The dS gravitational solutions described above admit a natural state interpretation in terms of the Euclidean gravitational path integral. For a given Dirichlet wall $\bar{\Sigma}^{i^{\mathrm{bra}}i^{\mathrm{ket}}}_c$, the dS gravitational path integral in the lower half of the Euclidean dS geometry produces a semiclassical dS microstate on the time reflection symmetric surface (Fig.~\ref{dS}). 
The Lorentzian dS gravitational solution provides a geometric realization of these semiclassical dS microstates and their subsequent time evolution. Moreover, since the Lorentzian dS gravitational solutions with different positive brane tension parameters $T_i$ share the same geometry between the dS horizon and the Dirichlet wall, they represent different semiclassical dS microstates of a common macrostate. These semiclassical dS microstates are distinguished by their geometry outside the dS horizon and are labeled by the tension parameter of the thin-shell brane. The set of these semiclassical dS microstates, denoted by $\{|\Psi_{T_i}\rangle\}$, forms an infinite family of states in the Hilbert space of the dS gravity theory. Below, we proceed to use the semiclassical Euclidean gravitational path integral to compute the nonperturbative overlaps between these semiclassical dS microstates.

\section{Nonperturbative overlaps between semiclassical dS microstates}
Consider a discrete and finite subset of the semiclassical dS microstates, $F_{\Omega}=\{|\Psi_{T_i}\rangle\,|\, i=1,2,\cdots,\Omega\}$, labeled by the brane tension parameter $T_i = iT_0$, where $T_0$ is a fixed tension scale. The normalized overlaps between these semiclassical dS microstates can be expressed as
\be
G_{ij}\equiv\frac{\langle\Psi_{T_{i}}|\Psi_{T_{j}}\rangle}{\sqrt{\langle\Psi_{T_{i}}|\Psi_{T_{i}}\rangle}\sqrt{\langle\Psi_{T_{j}}|\Psi_{T_{j}}\rangle}}=\frac{Z_{ij}}{\sqrt{Z_{ii}}\sqrt{Z_{jj}}}\ .
\ee
Here, in the second equality, the overlaps $\langle\Psi_{T_{i}}|\Psi_{T_{j}}\rangle$ are assumed to be evaluated via the semiclassical Euclidean gravitational path integral, and $Z_{ij}$ represent the dS gravitational path integral with a Dirichlet wall $\bar{\Sigma}_{c}^{i^{\mathrm{bra}}j^\mathrm{ket}}$. Moreover, the $n$-th moments of overlaps between the semiclassical dS microstates are expressed as
\be\label{nm}
G_{i_1i_2}G_{i_2i_3}\dots G_{i_ni_1}=\frac{Z_{i_1i_2}Z_{i_2i_3}\dots Z_{i_ni_1}}{Z_{i_1i_1}Z_{i_2i_2}\dots Z_{i_ni_n}}\ ,
\ee
where the numerator $Z_{i_1i_2}Z_{i_2i_3}\dots Z_{i_ni_1}$ represents the Euclidean dS gravitational path integral with $n$-Dirichlet walls $\bar{\Sigma}_{n,c}\equiv \bar{\Sigma}_{c}^{i_1^{\mathrm{bra}}i_{2}^\mathrm{ket}}\cup\bar{\Sigma}_{c}^{i_2^{\mathrm{bra}}i_{3}^\mathrm{ket}}\cup\cdots\cup\bar{\Sigma}_{c}^{i_n^{\mathrm{bra}}i_{1}^\mathrm{ket}}$. 

In the saddle-point approximation, the semiclassical Euclidean gravitational path integral with $n$ boundaries includes contributions from both disconnected saddles and connected wormhole saddles. In general, the path integral can be approximated by the dominant saddle. However, recent developments in black hole physics~\cite{Penington:2019kki,Almheiri:2019qdq,Almheiri:2020cfm} have revealed that nonperturbative corrections arising from the the sum over all wormhole saddles encode fine-grained information about the underlying quantum gravity theory. In what follows, we evaluate the nonperturbative wormhole contributions to the $n$-th moment of overlaps between the semiclassical dS microstates $\{|\Psi_{T_i}\rangle\}$. These contributions are crucial for extracting the dimension of the Hilbert space spanned by the semiclassical dS microstates.

The saddles that connect the $n$ Dirichlet walls $\bar{\Sigma}_{n,c}$ include both partially connected and fully connected wormhole configurations. Since the partially connected wormhole configurations are composed of fully connected wormholes with fewer boundaries, it is sufficient to construct the fully connected wormhole geometries in order to sum over all wormhole contributions. The fully connected $n$-boundary wormholes that contribute to the $n$-th moment of overlaps (\ref{nm}) are constructed as follows (the case of $n=4$ is shown in Fig.~\ref{4bdy}):
\begin{figure}
\centering
\includegraphics[width=0.38\linewidth]{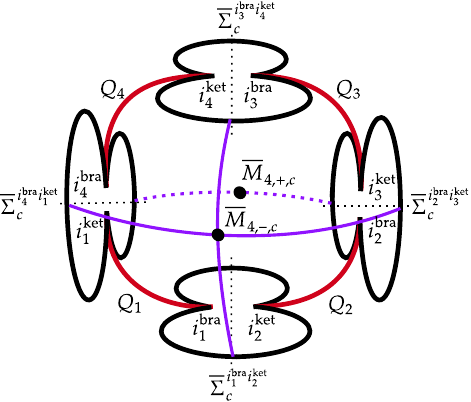}
\quad \quad \quad\quad\quad\quad\quad
\raisebox{0.3cm}{\includegraphics[width=0.35\linewidth]{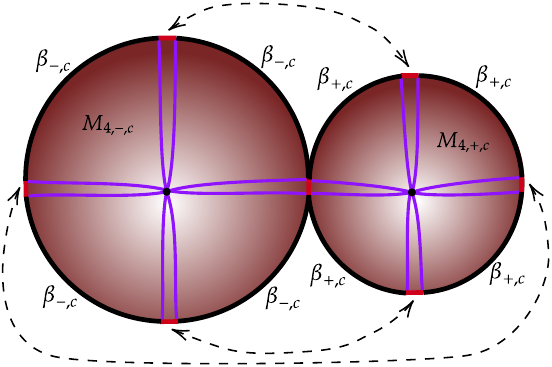}}
\caption{Left: The 4-boundary wormhole geometry is constructed by filling the Dirichlet walls
$\bar{\Sigma}_{4,\pm,c}\equiv \bar{\Sigma}_{\pm,c}^{i_1^{\mathrm{bra}}i_2^\mathrm{ket}}\cup \bar{\Sigma}_{\pm,c}^{i_2^{\mathrm{bra}}i_3^\mathrm{ket}}\cup \bar{\Sigma}_{\pm,c}^{i_3^{\mathrm{bra}}i_4^\mathrm{ket}}\cup \bar{\Sigma}_{\pm,c}^{i_4^{\mathrm{bra}}i_1^\mathrm{ket}}$
with two Euclidean pure dS geometries $\bar{M}_{4,\pm,c}$, which are glued together along four thin-shell branes $Q_k$ ($k=1,\dots,4$) connecting $i_k^{\mathrm{bra}}$ to $i_k^{\mathrm{ket}}$. Right: The fully connected 4-boundary wormhole that contributes to the fourth moment of overlaps between large tension semiclassical dS microstates is pinched into two nearly complete Euclidean pure dS geometries $M_{4,\pm,c}$. }
    \label{4bdy}
\end{figure}
\begin{enumerate}
    \item Fill the Dirichlet walls $\bar{\Sigma}_{n,\pm,c}$ with two Euclidean pure dS geometries $\bar{M}_{n,\pm,c}$, each having radius $l_{n,\pm}$ and a Dirichlet wall at $r=r_c$;

    \item Glue the two Euclidean pure dS geometries $\bar{M}_{n,\pm,c}$ using $n$ thin-shell branes $Q_k$  ($k=1,2,\dots,n$), each connecting the corresponding $i^{\mathrm{bra}}_k$ to $i^{\mathrm{ket}}_k$;
    
    \item Determine the trajectories of the thin-shell branes, and hence the resulting wormhole geometry, by requiring that all thin-shell branes $Q_k$ satisfy the Israel junction conditions.
\end{enumerate}

\begin{figure}
\centering
\includegraphics[scale=0.55]{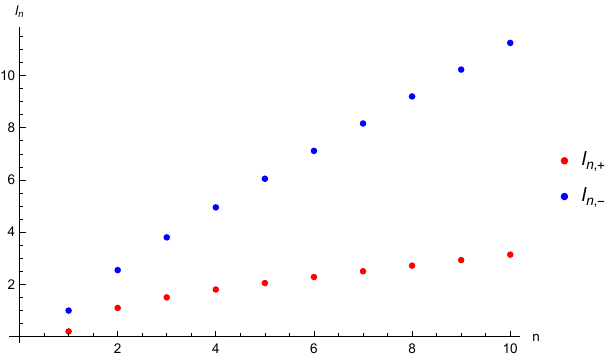}\quad\quad\quad\quad
\includegraphics[scale=0.55]{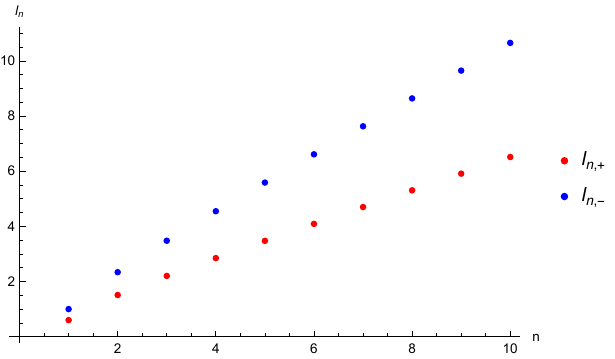}
\includegraphics[scale=0.55]{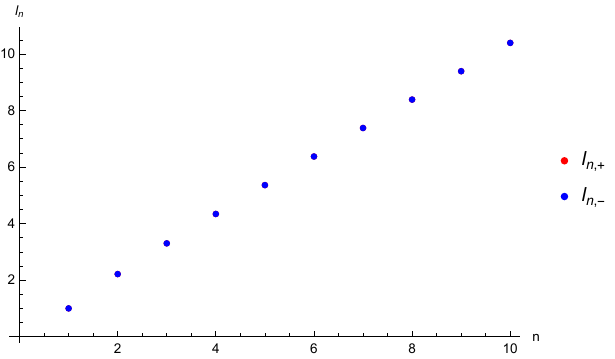}\quad\quad\quad\quad
\includegraphics[scale=0.55]{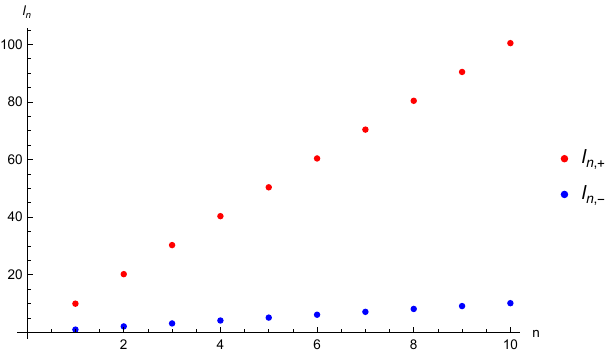}
    \caption{A numerical plot of dS radius of the fully connected $n$-boundary wormhole that contributes to $Z_{12}Z_{23}\dots Z_{n1}$ is shown. This diagram corresponds to $l_+=\frac{1}{5},\frac{3}{5},1,10$, and we have set $l_-=1$, $T_0=1$, $r_c=0.01$.}
    \label{wbp}
\end{figure}

The trajectories of the thin-shell branes $Q_k$ in $\bar{M}_{n,\pm,c}$ still satisfy the particle energy conservation equation (\ref{eomb}) but with $l_{\pm}$ and $T$ replaced by $l_{n,\pm}$ and $T_{i_k}$. In the fully connected $n$-boundary wormhole geometry, the sum of the proper lengths of the Dirichlet wall $\bar{\Sigma}_{n,\pm,c}$ and the boundary proper time elapsed by all the thin-shell branes $Q_k$ must be equal to the proper length of the Dirichlet wall of the full Euclidean pure dS geometries $M_{n,\pm,c}$, i.e. (see appendix~\ref{appB})
\be\label{nw}
\sum_{k=1}^{n}\left(\bar{\beta}_{\pm,c}^{i_k}+\sqrt{f_{n,\pm}(r_c)}\Delta \tau^{i_k}_{n,\pm}\right)=\beta_{n,\pm,c}\ ,
\ee
Here, $\bar{\beta}_{\pm,c}^{i_k}$ denotes the proper length of the Dirichlet wall $\Sigma_{\pm,c}^{i_k^{\mathrm{bra}}i_{k}^\mathrm{ket}}$, $\Delta \tau^{i_k}_{n,\pm}$ are Euclidean time elapsed by the thin-shell branes $Q_k$ in the wormhole geometry, and $\beta_{n,\pm,c}$ are proper length of the Dirichlet wall of the full Euclidean pure dS geometries $M_{n,\pm,c}$. Eq.~(\ref{nw}) provides a consistent condition to determine the dS radius $l_{n,\pm}$ of the $n$-boundary wormhole geometry. A numerical plot for $l_{n,\pm}$ from Eq.~(\ref{nw}) is shown in Fig.~\ref{wbp}.

To obtain the wormhole contributions to the $n$-th moment of overlaps in a controlled manner, we consider a set of large tension semiclassical dS microstates $\{|\Psi_{iT_0}\rangle\}$ with $T_0$ at least of order $\frac{1}{G_N}$. The normalized overlaps between the large tension states are $G_{ij}\approx \delta_{ij}$, since there is no smooth bulk saddle point solution that connects two states differing by a very large brane tension. Naively, the large tension states are orthogonal to each other. However, this is not the case once the nonperturbative wormhole corrections are included.

The fully connected $n$-boundary wormhole contribution to the $n$-th moment of overlaps between the large tension states is pinched into two nearly complete Euclidean pure dS geometries (Fig.~\ref{4bdy}).  To leading order in the large $T_0$ limit with $r\to0$, the Euclidean on-shell action of the pinched $n$-boundary wormhole is given by (see appendix~\ref{appC})
\be\label{acn}
I^E_n\approx I^E_{M_{n,+,c}}+I^E_{M_{n,-,c}}+\sum_{k=1}^{n}I^E_{Q_k}\ ,
\ee
where $I^E_{M_{n,\pm,c}}$ are the Euclidean on-shell actions of the full Euclidean pure dS geometries $M_{n,\pm,c}$. $I^E_{Q_k}$ is the Euclidean on-shell action of the thin-shell brane $Q_k$, which to leading order in the $\frac{1}{T_0}$ expansion does not depend on the bulk dS radius (Appendix~\ref{appC}). The partition function of the fully connected $n$-boundary wormhole saddle is
$Z_{i_1i_2}Z_{i_2i_3}\dots Z_{i_ni_1}|_{\mathrm{full\, conn.}}=e^{I_n^E}$. In particular, the normalization factor $Z_{i_ki_k}$ corresponds to the $n=1$ case. Then, the fully connected $n$-boundary wormhole saddle contributes to the $n$-th moments of overlaps (\ref{nm}) between the large tension semiclassical dS microstates as
\be\label{ug}
G_{i_1i_2}G_{i_2i_3}\dots G_{i_ni_1}|_{\mathrm{full\, conn.}}=\frac{Z\left(n\beta_{+,c}\right)Z\left(n\beta_{-,c}\right)}{Z\left(\beta_{+,c}\right)^nZ\left(\beta_{-,c}\right)^n}\ .
\ee
Here, $Z\left(n\beta_{\pm,c}\right) = e^{-I^E_{M_{n,\pm,c}}}$ is the partition function of the Euclidean pure dS geometry $M_{n,\pm,c}$, and the contributions from the thin-shell branes in (\ref{ug}) cancel out.

So far, the semiclassical dS microstates that we have considered are states in a canonical ensemble with temperature $\beta_{\pm,c}$. To properly account for the dimension of the Hilbert space spanned by the semiclassical dS microstates, we need to work in a microcanonical ensemble, which only involves a finite number of states within a given energy window. Consider a microcanonical ensemble constructed from the large tension states $|\Psi_{iT_{0}}^{E_c}\rangle$ within the energy window $[E_{c}, E_{c} + \delta E_{c}]$, where $E_c\equiv E_{+,c}+E_{-,c}$. In general, the canonical partition function $Z\left(\beta_{\pm,c}\right)$ is related to microcanonical degeneracy $\rho\left(E_{c,\pm}\right)$ of the microcanonical ensemble as~\cite{Balasubramanian:2022gmo}
\be
Z\left(\beta_{\pm,c}\right)=\int \mathrm{d}E_{\pm,c}\;\rho\left(E_{\pm,c}\right)e^{-\beta_{\pm,c} E_{\pm,c}}\ .
\ee
In the corresponding microcanonical window, the microcanonical entropy and the microcanonical version of $Z\left(\beta_{\pm,c}\right)$ are defined as~\cite{Penington:2019kki}
\be\label{miz}
e^{\mathbf{S}_{\pm}}\equiv\rho\left(E_{\pm,c}\right)\Delta E_{\pm,c}\ ,\quad \mathbf{Z}_{\pm}\equiv \rho\left(E_{\pm,c}\right)e^{-\beta_{\pm,c}E_{\pm,c}}\Delta E_{\pm,c}\ ,
\ee
where $\mathbf{S}_{\pm}=\frac{A_{\pm}}{4G}$ denote the microcanonical entropies associated with the dS horizons. Similarly, the microcanonical version of $Z\left(n\beta_{\pm,c}\right)$ is
\be\label{mizn}
\mathbf{Z}_{n,\pm}\equiv \rho\left(E_{\pm,c}\right)e^{-n\beta_{\pm,c}E_{\pm,c}}\Delta E_{\pm,c}\ .
\ee
Using Eqs.~(\ref{miz}) and (\ref{mizn}), we find that the fully connected $n$-boundary wormhole saddle contributes to the $n$-th moment of overlaps between the microcanonical large tension semiclassical dS microstates as
\be\label{micc}
\mathbf{G}_{i_1i_2}\mathbf{G}_{i_2i_3}\dots \mathbf{G}_{i_ni_1}|_{\mathrm{full\, conn.}}=e^{\left(1-n\right)\left(\mathbf{S}_{+}+\mathbf{S}_{-}\right)}\ .
\ee

\section{Microstates counting of dS entropy}
Now we count the dimension of the Hilbert space spanned by the microcanonical semiclassical dS microstates $\mathbf{F}_{\Omega}\equiv\{|\Psi^{E_c}_{T_i}\rangle\,|\, i=1,2,\cdots,\Omega\}$. 
The number of linearly independent states in this set is given by the rank of its overlap matrix $\mathbf{G}$. Since the overlap matrix is a Hermitian positive semidefinite matrix, its rank is determined by the number of its positive eigenvalues, which can be extracted from the trace of its resolvent matrix~\cite{Penington:2019kki,Balasubramanian:2022gmo} 
\be\label{tresol}
R\left(\lambda\right)\equiv  \sum_{i=1}^{\Omega}\left(\frac{1}{\lambda\,  \mathbb{I}-\mathbf{G}}\right)_{ii}=\frac{\Omega}{\lambda}+\sum_{n=1}^{\infty}\frac{1}{\lambda^{n+1}}\sum_{i=1}^{\Omega}\left(\mathbf{G}^n\right)_{ii}\ ,
\ee
where $\left(\mathbf{G}^n\right)_{ii}=\sum_{i_2,\dots,i_n=1}^{\Omega}\mathbf{G}_{ii_2}\mathbf{G}_{i_2i_3}\dots \mathbf{G}_{i_ni}$. More precisely, the density of eigenvalues of the overlap matrix is given by the discontinuity of $R(\lambda)$ across the real axis as
\be\label{Ds}
D\left(\lambda\right)=\lim_{\epsilon\rightarrow0}\frac{1}{2\pi \mathrm{i}}\left(R\left(\lambda-\mathrm{i}\epsilon\right)-R\left(\lambda+\mathrm{i} \epsilon\right)\right)\ .
\ee
The number of positive eigenvalues of the overlap matrix is given by
\be\label{dim}
d_{\Omega}=\lim_{\epsilon\rightarrow 0^{+}}\int_{\epsilon}^{\infty}\mathrm{d}\lambda \;D\left(\lambda\right)\ .
\ee

Using the semiclassical Euclidean gravitational path integral representation, the $n$-th moments of overlaps $\mathbf{G}_{ii_2}\mathbf{G}_{i_2i_3}\dots \mathbf{G}_{i_ni}$ appearing in Eq.~(\ref{tresol}) are given by the sum over all of the $n$-boundary planar disconnected and connected saddle point geometries. In terms of the saddle point geometric configurations, the sum in Eq.~(\ref{tresol}) can be represented as
\be
\bal
\rt{\,}&=\rtz{\,}+\rto{\,}+\rttd{\,}+\rttc+\dots\\
&=\rtz{\,}+\srto{\,}+\srtt{\,}+\dots\ .
\eal
\ee
where each straight dashed line comes with a factor of $\frac{1}{\lambda}$, and each curved dashed line represents a sum over brane indices ranging from $1$ to $\Omega$. In the second equality, the sums over all planar geometries are reorganized into a sum over all fully connected geometries, while all other terms are resummed into $R\left(\lambda\right)$. This resummation leads to a Schwinger-Dyson equation for the trace of the resolvent matrix
\be\label{sd}
R\left(\lambda\right)=\frac{\Omega}{\lambda}+\frac{1}{\lambda}\sum_{n=1}^{\infty}\frac{R\left(\lambda\right)^n }{\Omega^n}\sum_{i,i_2,\dots,i_n=1}^{\Omega}\mathbf{G}_{ii_2}\mathbf{G}_{i_2i_3}\dots \mathbf{G}_{i_ni}|_{\mathrm{full\ conn.}}
\ .
\ee
Substituting the expression from Eq.~(\ref{micc}) into Eq.~(\ref{sd}), we obtain a quadratic equation for $R(\lambda)$, i.e., 
\be
R\left(\lambda\right)=\frac{\Omega}{\lambda}+\frac{1}{\lambda}\frac{R\left(\lambda\right) e^{\mathbf{S}_{+}+\mathbf{S}_{-}}}{e^{\mathbf{S}_{+}+\mathbf{S}_{-}}-R\left(\lambda\right)}\ .
\ee
By solving this equation and using Eq.~(\ref{Ds}), the density of eigenvalues of the overlap matrix can be obtained as
\be\label{dss}
\bal
D\left(\lambda\right)=&\frac{e^{\mathbf{S}_{+}+\mathbf{S}_{-}}}{2\pi\lambda}\sqrt{\left(\lambda-\left(1-\Omega^{\frac{1}{2}}e^{-\frac{\mathbf{S}_{+}+\mathbf{S}_{-}}{2}}\right)^2\right)\left(\left(1+\Omega^{\frac{1}{2}}e^{-\frac{\mathbf{S}_{+}+\mathbf{S}_{-}}{2}}\right)^2-\lambda\right)}\\
&+\delta\left(\lambda\right)\left(\Omega-e^{\mathbf{S}_{+}+\mathbf{S}_{-}}\right)\theta\left(\Omega-e^{\mathbf{S}_{+}+\mathbf{S}_{-}}\right)\ .
\eal
\ee

The rank of the overlap matrix is further determined by $D(\lambda)$ via Eq.~(\ref{dim}). Since the density of eigenvalues in Eq.~(\ref{dss}) is separated into a continuous part and a singular part, the integrals of these two parts count the numbers of zero and positive eigenvalues, respectively. By integrating these two parts separately, we find that when $\Omega<e^{\mathbf{S}_{+}+\mathbf{S}_{-}}$, the overlap matrix has no zero eigenvalues, and its rank is given by $\Omega$. When $\Omega>e^{\mathbf{S}_{+}+\mathbf{S}_{-}}$, the overlap matrix has $\Omega-e^{\mathbf{S}_{+}+\mathbf{S}_{-}}$ zero eigenvalues, and its rank is given by $e^{\mathbf{S}_{+}+\mathbf{S}_{-}}$. This implies that the number of orthogonal states within the infinite family of semiclassical dS microstates $\{|\Psi_{T_{i}}^{E_c}\rangle\,|\,i=1,2,\dots\}$ is $e^{\mathbf{S}_{+}+\mathbf{S}_{-}}$, which precisely matches the Gibbons-Hawking entropy associated with the left and right static patches of the dS spacetime. Therefore, these semiclassical dS microstates provide a microscopic accounting for the Gibbons–Hawking entropy of de Sitter spacetime.

\section{Conclusions and Discussions}
In this work, we provide a gravitational state-counting derivation for Gibbons-Hawking entropy of dS spacetime. This was achieved by first constructing an infinite family of semiclassical dS microstates, realized as backreacted geometries of dS spacetime with a constant tension thin-shell brane located outside the dS horizon. Then we extract nonperturbative overlaps between the semiclassical dS microstates by considering the wormhole contributions in the semiclassical Euclidean gravitational path integral. We find that the small and universal overlaps between the semiclassical dS microstates lead them to span a Hilbert space with dimension equal to the exponential of the Gibbons-Hawking entropy of the dS spacetime, and thus explain the microscopic origin of dS entropy. 

Our construction provides new insights into the nonperturbative effects of quantum gravity in cosmological spacetimes and raises several directions for future research. First, the semiclassical dS microstate geometry, which involves both a timelike Dirichlet wall and a thin-shell brane, may be interpreted as the bulk dual of a holographic interface CFT~\cite{Karch:2000gx, Aharony:2003qf, Clark:2004sb, Takayanagi:2011zk} with a $T\bar{T}\,\,  (+\Lambda_2)$ deformation~\cite{Gorbenko:2018oov, Lewkowycz:2019xse,Coleman:2021nor,Wang:2024jem}. In addition, the physical properties of the semiclassical dS microstate geometry, such as its phase structure~\cite{Bachas:2021fqo}, energy transmission behavior~\cite{Bachas:2020yxv,Bachas:2022etu}, and quantum information characteristics~\cite{Karch:2024udk}, merit further study. Moreover, our construction can also be applied to account for quantum corrections to dS entropy~\cite{Anninos:2020hfj}, as in the black hole case~\cite{Climent:2024trz}. Last but not least, our construction provides an intriguing avenue to explore in more detail the Hilbert space structure~\cite{Balasubramanian:2024yxk,Balasubramanian:2025zey} and other fine-grained properties of a closed universe, potentially extending even to scenarios involving an observer~\cite{Chandrasekaran:2022cip, Abdalla:2025gzn, Harlow:2025pvj}.

\appendix

\section{Spherical thin-shell brane in dS spacetime}\label{appA}
We derive the equation of motion for a time-reversal symmetric spherical thin-shell brane in dS spacetime beginning with the Euclidean theory. The bulk geometries on either side of the thin-shell brane are taken to be Euclidean pure dS geometries with a Dirichlet wall located at $r=r_{c}$ (Fig.~{\ref{dS}})
\be\label{edd}
\mathrm{d}s_{\pm}^2=f(r)\mathrm{d}\tau_{\pm}^2+\frac{1}{f(r)}\mathrm{d}r^2+r^2\left(\mathrm{d}\theta^2+\sin^2\theta \mathrm{d}\Omega_{d-2}^2\right)\ ,\quad f(r)=1-\frac{r^2}{l_{\pm}^2}\ ,\quad r\geq r_c\ .
\ee
The spherical thin-shell brane embedded in these two bulk geometries is parameterized as $(\tau_{\pm}(\tau_b),R(\tau_b))$, where $\tau_b$ denotes the Euclidean proper time of the brane. The configuration of the thin-shell brane is determined by the Israel junction conditions
\begin{align}
h_{+,ab}&=h_{-,ab} \ ,\label{nbc1}\\
K_{+,ab}+K_{-,ab}&=-Th_{ab}\ .\label{nbc2}
\end{align}

The first Israel junction condition (\ref{nbc1}) requires that the induced metrics from both sides of the spherical thin-shell brane coincide, i.e.,
\be\label{I}
f_\pm(R) \dot{\tau}_{\pm}^2+\frac{1}{f_{\pm}(R)}\dot{R}^2=1\ ,
\ee
where the dot represents the derivative with respect to $\tau_b$. The second Israel junction condition (\ref{nbc2}) demands that the discontinuity of the extrinsic curvature across the brane is balanced by the stress-energy tensor localized on the brane. The extrinsic curvature of the thin-shell brane is defined as $K_{\pm,ab}=\frac{\partial x^{\mu}}{\partial y^a}\frac{\partial x^{\nu}}{\partial y^b}\nabla_{\mu}n_{\pm,\nu}$, where the outward-pointing normal vector of the thin-shell brane is $n_{\pm,\nu}=(-\dot{R},\dot{\tau}_{\pm})$. In particular, the $\theta\theta$ component of the extrinsic curvature is calculated as
\be
K_{\pm,\theta\theta}=\nabla_{\theta}n_{\pm,\theta}=-\Gamma^{r}_{\theta\theta}n_{\pm,r}=Rf_{\pm}(R)\dot{\tau}_{\pm}\ .
\ee
Therefore, the $\theta\theta$ component of the second Israel junction condition can be expressed as
\be\label{II}
f_{+}(R)\dot{\tau}_{+}+f_{-}(R)\dot{\tau}_{-}=-TR \ .
\ee
We can check that the other components of the second Israel junction condition are not independent of (\ref{II}).
Combining Eq.~(\ref{I}) and Eq.~(\ref{II}), we find that the radius of the thin-shell brane satisfies the equation of motion of a particle with zero total energy,
\be\label{rd}
\dot{R}^2+V^E_{\mathrm{eff}}(R)=0\ ,\quad V^E_{\mathrm{eff}}(R)=-f_+(R)+\left(\frac{f_-(R)-f_+(R)-T^2R^2}{2T R}\right)^2.
\ee

Moreover, using Eq.~(\ref{I}) and Eq.~(\ref{rd}), we obtain the equation of motion of the thin-shell brane expressed in terms of the static coordinates as
\be\label{eomst}
\frac{\mathrm{d}\tau_{\pm}}{\mathrm{d}R}=\frac{\dot{\tau}_\pm}{\dot R}=\frac{1}{f_{\pm}(R)}\sqrt{\frac{f_{\pm}(R)+V^E_{\mathrm{eff}}(R)}{-V^E_{\mathrm{eff}}(R)}}\quad\text{for}\quad \tau_b \ge 0\ .
\ee
where $l_b$ is the radius of the thin-shell brane at $\tau_b=0$, which is determined by $V^E_{\mathrm{eff}}(R)=0$.
The Euclidean time elapsed by the thin-shell brane is
\be\label{dt}
\Delta \tau_{\pm}=2\int_{r_c}^{l_b}\;\mathrm{d}R\frac{1}{f_{\pm}(R)}\sqrt{\frac{f_{\pm}(R)+V^E_{\mathrm{eff}}(R)}{-V^E_{\mathrm{eff}}(R)}}\ .
\ee
For a given bulk geometry (\ref{edd}) and a brane tension parameter $T$, the proper length of the Dirichlet wall in the glued Euclidean dS geometry is
\be\label{bb}
\bar{\beta}_{\pm,c}=\beta_{\pm,c}-\sqrt{f_{\pm}(r_c)}\Delta\tau_{\pm}\ .
\ee

The time-reversal-symmetric Euclidean dS geometry, determined by Eqs.~(\ref{rd}) and (\ref{eomst}), can be analytically continued to a Lorentzian dS spacetime along the time-reversal-symmetric surface. In the Lorentzian signature, the radius of the thin-shell brane, $R(t_b)$, satisfies the equation of motion for a particle with zero total energy in the effective potential $V_{\mathrm{eff}}(R)=-V^E_{\mathrm{eff}}(R)$.

\section{Multi-boundary dS wormhole}\label{appB}
The fully connected $n$-boundary wormhole geometries that contribute to the $n$-th moment of overlaps $G_{i_1i_2}G_{i_2i_3}\dots G_{i_ni_1}$ are constructed by gluing the Euclidean pure dS geometries $\bar{M}_{n,\pm,c}$ 
\be\label{men}
\mathrm{d}s_{n,\pm}^2=f_{n,\pm}(r)\mathrm{d}\tau_{\pm}^2+\frac{1}{f_{n,\pm}(r)}\mathrm{d}r^2+r^2 \mathrm{d}\Omega_{d-1}^2\ ,\quad f_{n,\pm}(r) \equiv 1-\frac{r^2}{l_{n,\pm}^2},\quad r\geq r_c
\ee
along the $n$ thin-shell branes $Q_k$, each of which connects $i^{\mathrm{bra}}_k$ to $i^{\mathrm{ket}}_k$ (Fig.~\ref{4bdy}). The equations of motion for each thin-shell brane $Q_k$ and the Euclidean time it elapses in the $n$-boundary wormhole geometry are still given by Eqs.~(\ref{rd}) and Eq.~({\ref{dt}}), except that $l_{\pm}$ and $T$ are replaced by $l_{n,\pm}$ and $T_{i_k}$.

The Dirichlet walls of the $n$-boundary wormhole geometry are given by $\Sigma_{n,\pm,c}\equiv\Sigma_{\pm,c}^{i_1^{\mathrm{bra}}i_2^\mathrm{ket}}\cup\Sigma_{\pm,c}^{i_2^{\mathrm{bra}}i_3^\mathrm{ket}}\cup \dots\cup\Sigma_{\pm,c}^{i_n^{\mathrm{bra}}i_1^\mathrm{ket}}$. The proper length of the  component $\Sigma_{\pm,c}^{i_k^{\mathrm{bra}}i_{k+1}^\mathrm{ket}}$ in  the Dirichlet walls $\Sigma_{n,\pm,c}$ is $\bar{\beta}_{\pm,c}^{i_{k}^{\mathrm{bra}}i_{k+1}^{\mathrm{ket}}}=\frac{\bar{\beta}_{\pm,c}^{i_{k}}+\bar{\beta}_{\pm,c}^{i_{k+1}}}{2}$, where $\bar{\beta}_{\pm,c}^{i_{k}}$ is given by Eq.~(\ref{bb}). Then the total length of the Dirichlet wall of the $n$-boundary wormhole is given by $\bar{\beta}_{n,\pm,c}=\sum_{k=1}^{n}\bar{\beta}_{\pm,c}^{i_{k}}$. The sum of the proper length of the Dirichlet walls of the $n$-boundary wormhole geometry and the total proper boundary time elapsed by all the thin-shell branes is required to be consistent with the proper length of the Dirichlet wall of the full Euclidean pure dS geometries $M_{n,\pm,c}$ , i.e.
\be
\sum_{k=1}^{n}\left(\bar{\beta}_{\pm,c}^{i_k}+\sqrt{f_{n,\pm}(r_c)}\Delta \tau^{i_k}_{n,\pm}\right)=\beta_{n,\pm,c}\ ,
\ee
which provides an equation to determine the dS radius of the $n$-boundary wormhole geometry.

\section{Euclidean on-shell action of the multi-boundary dS wormhole}\label{appC}
We begin by computing the Euclidean on-shell action of a semiclassical dS microstate geometry with a single thin-shell brane $Q$ characterized by a tension parameter $T$. The corresponding saddle-point geometry can be divided into five parts as
\begin{equation}
    \FB=\LL+\LR+\BR+\RL+\RR\ .
\end{equation}
The $n$-boundary wormhole geometry can be decomposed in the same manner, and these five parts constitute the fundamental building blocks for computing its on-shell action. The geometrical quantities of these five parts are
\be
\bal
R_{\pm}-2\Lambda_{\pm}&=\frac{d(d+1)}{l_{\pm}^2}-\frac{d(d-1)}{l_{\pm}^2}=\frac{2d}{l_{\pm}^2}\ ,\\
\quad K_{\pm,c}&=\sqrt{f_{\pm}(r_c)}\frac{\mathrm{d}}{\mathrm{d}r}\left( \sqrt{f_{\pm}(r)}r^{d-1}\right)|_{r=r_c}=\frac{dr_c^d}{l_{\pm}^2}-(d-1)r_c^{d-2}\ , \\
K_{+,Q}-K_{-,Q}&=-dT\ .
\eal
\ee

The on-shell action of the bulk parts enclosed by the Dirichlet walls $\Sigma_{\pm,c}$ is
\be
\bal
I^E_{\RRs/\LLs}&=-\frac{1}{16\pi G_N}\int_{\RRs/\LLs}\sqrt{g}\left(R-2\Lambda\right)-\frac{1}{8\pi G_N}\int_{\bar{\Sigma}_{+,c}/\bar{\Sigma}_{-,c}}\sqrt{\gamma}K_{\pm,c}\\
&=-\frac{1}{16\pi G_N}\frac{2d}{l_{\pm}^2}V_{\mathrm{S}^{d-1}}\bar{\beta}_{\pm,c}\int_{r_c}^{l_\pm}\mathrm{d}r\,r^{d-1}+\frac{1}{8\pi G_N}V_{\mathrm{S}^{d-1}}\bar{\beta}_{\pm,c}K_{\pm,c}\\
&=\ \frac{\left(d-1\right)r_c^{d-2}\left(l_{\pm}^2-r_c^2\right)-l_{\pm}^d}{4\pi G_Nl_{\pm}}V_{\mathrm{S}^{d-1}}\left(\pi-\arctan \sqrt{\frac{l_{b}^2-r_c^2}{l_{\pm}^2-l_b^2}}\right)\ .
\eal
\ee
The on-shell action of the bulk parts enclosed by the thin-shell brane $Q$ is
\be
\bal
I^E_{\RLs/\LRs}&=-\frac{1}{16\pi G_N}\int_{\RLs/\LRs}\sqrt{g}\left(R-2\Lambda\right)=-\frac{d}{8\pi G}\frac{1}{l_{\pm}^2}\mathrm{Vol}[\RL/\LR]\\
&=-\frac{d}{8\pi G_N}\frac{1}{l_{\pm}^2}V_{\mathrm{S}^{d-1}}\int_{-\frac{\Delta \tau_\pm}{2}}^{\frac{\Delta \tau_\pm}{2}}\mathrm{d}\tau_{\pm}\int_{r(\tau_{\pm})}^{l_{\pm}}\mathrm{d}r\,r^{d-1}\\
&=\frac{l_{\pm}^{d-1}V_{\mathrm{S}^{d-1}}}{4\pi G_N}\left(\frac{l_b^d}{l_{\pm}^d}\sqrt{\frac{l_{b}^2-r_c^2}{l_{\pm}^2-l_b^2}}F_1\left[\frac{1}{2};-\frac{d}{2},1;\frac{3}{2};1-\frac{r_c^2}{l_b^2},-\frac{l_{b}^2-r_c^2}{l_{\pm}^2-l_b^2}\right]-\arctan \sqrt{\frac{l_{b}^2-r_c^2}{l_{\pm}^2-l_b^2}}\right)\ ,
\eal
\ee
where $F_1[a;b_1,b_2;c;x,y]$ is Appell function of the first kind. The on-shell action of the thin-shell brane is
\be\label{brt}
\bal
I^E_{\BRs}&=-\frac{1}{8\pi G_N}\int_{\BRs}\sqrt{h}\left(K_{Q,+}-K_{Q,-}+\left(d-1\right)T\right)=\frac{T}{8\pi G}\mathrm{Vol}[\ \BR\ ]\\
&=\frac{T}{8\pi G_N}V_{\mathrm{S}^{d-1}}\int_{-\tau^c_{b}}^{\tau^c_{b}}\mathrm{d}\tau_{b}\ R(\tau_b)^{d-1}\\
&=\frac{T V_{\mathrm{S}^{d-1}}}{8\pi G_N}
\left(\frac{l_b^d\sqrt{\pi}\Gamma(\frac{d}{2})}{\Gamma(\frac{d+1}{2})}-\frac{2r_c^d}{d} \,_2F_1\left[\frac{1}{2},\frac{d}{2};1+\frac{d}{2};\frac{r_c^2}{l_b^2}\right]\right)\ ,
\eal
\ee
where $\tau_{b}^c$ represents the Euclidean  proper time of the brane at the intersection point between the thin-shell brane and the Dirichlet wall, and $_2F_1[a,b;c;z]$ is Gauss hypergeometric function. In particular, for $d=2$, the on-shell action of the bulk geometries can be expressed analytically as
\be
\bal
I_{\RRs/\LLs}&=-\frac{r_c^2}{2G_Nl_{\pm}}\left(\pi-\arctan \sqrt{\frac{l_{b}^2-r_c^2}{l_{\pm}^2-l_b^2}}\right)\ ,\\
I_{\RLs/\LRs}&=-\frac{1}{G_Nl_{\pm}}\sqrt{\left(l_{\pm}^2-l_b^2\right)\left(l_b^2-r_c^2\right)}\ ,\\
I_{\BRs}&=\frac{Tl_b}{2G_N}\sqrt{l_b^2-r_c^2}\ .
\eal
\ee

For large tension states with $T$ at least of order $\frac{1}{G_N}$ and with $r_c\to 0$, the on-shell action of these five parts can be expanded with respect to $\frac{1}{T}$ as
\begin{align}
I^E_{\RRs/\LLs}&=-\frac{l_{\pm}^{d-1} V_{\mathrm{S}^{d-1}}}{4}\frac{1}{G_N}+\frac{l_{\pm}^{d-2}V_{\mathrm{S}^{d-1}}}{2\pi}\frac{1}{G_N}\frac{1}{T}+\frac{1}{G_{N}}O(\frac{1}{T^2})\ ,\\
I^E_{\RLs/\LRs}&=-\frac{l_{\pm}^{d-2}V_{\mathrm{S}^{d-1}}}{2\pi }\frac{1}{G_N}\frac{1}{T}+\frac{1}{G_N}O(\frac{1}{T^2})\ ,\\
I^E_{\BRs}&=\frac{2^{d-3}\Gamma(\frac{d}{2})V_{\mathrm{S}^{d-1}}}{\sqrt{\pi}\Gamma(\frac{d+1}{2})}\frac{1}{G_N}\frac{1}{T^{d-1}}+\frac{1}{G_N}O(\frac{1}{T^d})\ .
\end{align}
We find that the tension dependence in the bulk parts is always canceled with each other, so the total on-shell action of the semiclassical dS microstates is given by
\be
\bal
I^E_{\FBs}&=-\frac{l_{+}^{d-1} V_{\mathrm{S}^{d-1}}}{4G_N}-\frac{l_{-}^{d-1} V_{\mathrm{S}^{d-1}}}{4G_N}+\frac{2^{d-3}\Gamma(\frac{d}{2})V_{\mathrm{S}^{d-1}}}{\sqrt{\pi}\Gamma(\frac{d+1}{2})}\frac{1}{G_N}\frac{1}{T^{d-1}}+\frac{1}{G_N}O(\frac{1}{T^d})\\
&=I^E_{M_{+}}+I^E_{M_{-}}+I_Q^{E}\ .
\eal
\ee
Using this building block, the on-shell action of the $n$-boundary wormhole geometry is obtained as
\be
I_n=I^E_{M_{n,+}}+I^E_{M_{n,-}}+\sum_{k=1}^{n}I_{Q_k}^E\ .
\ee
Note that the leading term of the on-shell action of the thin-shell brane does not depend on the bulk dS radius. This leads to a cancellation of the thin-shell brane contribution to the normalized $n$-th moments of overlaps for the large tension states.

\bibliographystyle{JHEP}
\bibliography{dSentropy}

\end{document}